\input{aipcheck}

\documentclass[
    ,final            
  ]
  {aipproc}

\layoutstyle{6x9}

\begin{document}

\title{Physical Classification of Galaxies with MOPED/VESPA}


\author{Raul Jimenez}{
  address={ICREA \& ICE(CSIC)-IEEC, UAB campus, Bellaterra 08193, Spain}
  ,altaddress={Dept. of Astrophysical Sciences, Princeton University, Peyton Hall, Princeton 08544, USA} 
}

\author{Alan F. Heavens}{
  address={SUPA Institute for Astronomy, ROE Blackford Hill, Edinburgh EH9-3HJ, UK}
}

\author{Ben Panter}{
  address={SUPA Institute for Astronomy, ROE Blackford Hill, Edinburgh EH9-3HJ, UK}
}

\author{Rita Tojeiro}{
  address={SUPA Institute for Astronomy, ROE Blackford Hill, Edinburgh EH9-3HJ, UK}
}

\begin{abstract}
 The availability of high-quality spectra for a large number of galaxies in the SDSS survey allows for 
 a more sophisticated extraction of information about their stellar populations than, e.g., the luminosity 
 weighted age. Indeed, sophisticated and robust techniques to fully analyze galaxy spectra have now reached enough maturity as to 
 trust their results and findings. By reconstructing the star formation and metallicity history of galaxies from the SDSS fossil record and analyzing how it relates to its environment, we have learned how to classify galaxies: to first order the evolution of a galaxy is determined by its present stellar mass, which in turn seems to be governed by the merger rate of dark halos.
\end{abstract}

\maketitle


\section{Introduction}

There has been substantial recent progress in the development of 
methods which determine the star formation and chemical composition 
histories of galaxies from the integrated spectra of their stellar 
populations.  The traditional approach has been to determine the instantaneous star 
formation rate or metallicity from certain features in the spectrum 
of a galaxy.  However, several recent algorithms 
\cite{HJL00, Sodre05, Mathis06, Ocvirk06}, have been developed 
which use the entire spectrum to infer the entire star formation 
history and the evolution of the chemical composition of the object, 
rather than simply the instantaneous values of these quantities.  
MOPED \cite{HJL00} and VESPA \cite{Rita}, have been used to determine the 
star formation histories and metallicities of galaxies drawn from 
the Sloan Digital Sky Survey 
\cite{PHJ03,HPJD04,PHJ04,JPHV05,PJHC06a}.

There has also been significant progress in quantifying how 
galaxies are distributed on large scales, and using this to 
constrain cosmological parameters \citep{2dFPk05, SDSSPk05}.  
In such analyses of galaxy clustering, it is common to treat 
galaxies as points, ignoring the fact that galaxies have different 
luminosities, colors, masses, star formation histories, metallicities, 
etc.  However, as a result of improvements in detector technology, 
and in the algorithms such as MOPED and VESPA with which the new data is 
analyzed, many such galaxy attributes are now sufficiently reliably 
measured that one can use them as weights when studying the 
clustering of galaxies.  Thus, one can now study the clustering of 
luminosity, color, star formation rate, etc.  Mark statistics 
\citep{stoyan2,BK00} provide a useful framework for describing 
attribute-weighted clustering.  Moreover, they provide sensitive 
probes of how the properties of galaxies correlate with their 
environments. In this respect, mark statistics 
provide a useful link between the large-scale structures which 
galaxies trace, and the properties of those galaxies.  

Further, it is reasonable to expect that the spin ($\lambda$) of a dark matter halo will influence the final properties of baryonic matter in the galaxy. For a simple example, consider a disk galaxy and assume baryons settle into the disk with no loss of angular momentum. $\lambda$ measures the degree to which rotation contributes to supporting the galaxy against collapse, between negligibly ($\lambda=0$) and completely ($\lambda \sim 1$). Higher $\lambda$ disks are more rotationally supported and will therefore be less dense. When coupled with a star formation law dependent on density, higher $\lambda$ further implies less efficient star forming systems.

Using analytical models, \cite{Dalcanton97}, \cite{JHHP97}, \cite{vdB98}, and \cite{mo} showed in detail how a distribution in the values of the halo spin parameter could lead to
significant differences in the star formation efficiency of the disk, thus shaping the history of the galaxy,
and therefore the morphological type, even leading to possible cases of
dark galaxies \citep{Dalcanton97,JHHP97,VOJ02}, where star formation in the disk has been completely prevented.

Thus, one expects that mass, environment and spin will all influence the evolution of galaxies. Below, we show quantitatively how each of these affects the evolution of galaxies and conclude that mass alone dominates the evolution of galaxies.

\section{Downsizing: classifying galaxies by mass}

\begin{figure}[ht]
\includegraphics[width=0.8\textwidth]{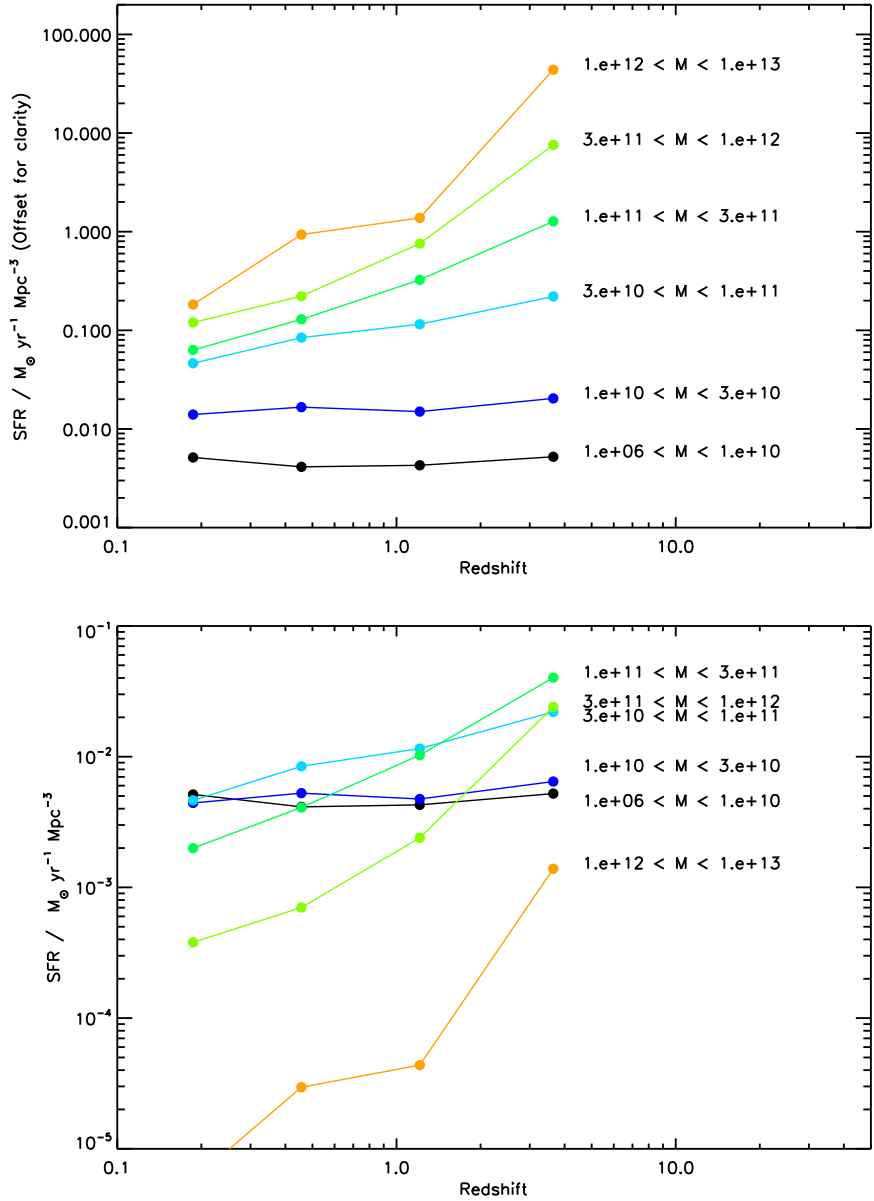}
\caption{The star formation rate of galaxies of different masses.
These plots show the contribution to the overall star formation rate
in the universe from galaxies with different masses over the
redshift range we consider reliable. In the upper panel the SFR has
been offset to enable easier comparison of the curves, in the lower
there has been no offset applied. It is clear that the more massive
systems formed their stars earlier. Figure from [10].}
\label{fig:downsizing}
\end{figure}

One of the results of \cite{HPJD04,PJHC06a} was the finding of `downsizing'
from the SDSS fossil record. With the new BC03 models at higher
resolution they found that the evidence for downsizing is just as
clear. In Fig. \ref{fig:downsizing} we show the cosmic star
formation rate for galaxies split into different stellar mass
ranges. A clear signature of `downsizing' is seen: the stars ending
up in today's highest-mass galaxies formed early, and show
negligible recent star formation, while the lower-mass galaxies
continue with star formation until the present day. The lower,
non-offset plot can be used to determine for a given redshift which
galaxies dominate the star formation rate. Clearly, galaxies could be classified by their present 
stellar mass as this would tell us what star formation they will have during their entire history. If the star formation is associated to the final stellar mass of the galaxy, is the mass determined by environment? or is star formation influenced by environment? What is the role of merging in shaping the star formation history of galaxies? We answer these questions below.

\section{How do galaxies relate to the environment?}

\begin{figure}[ht]
 \centering
 \includegraphics[width=0.9\columnwidth]{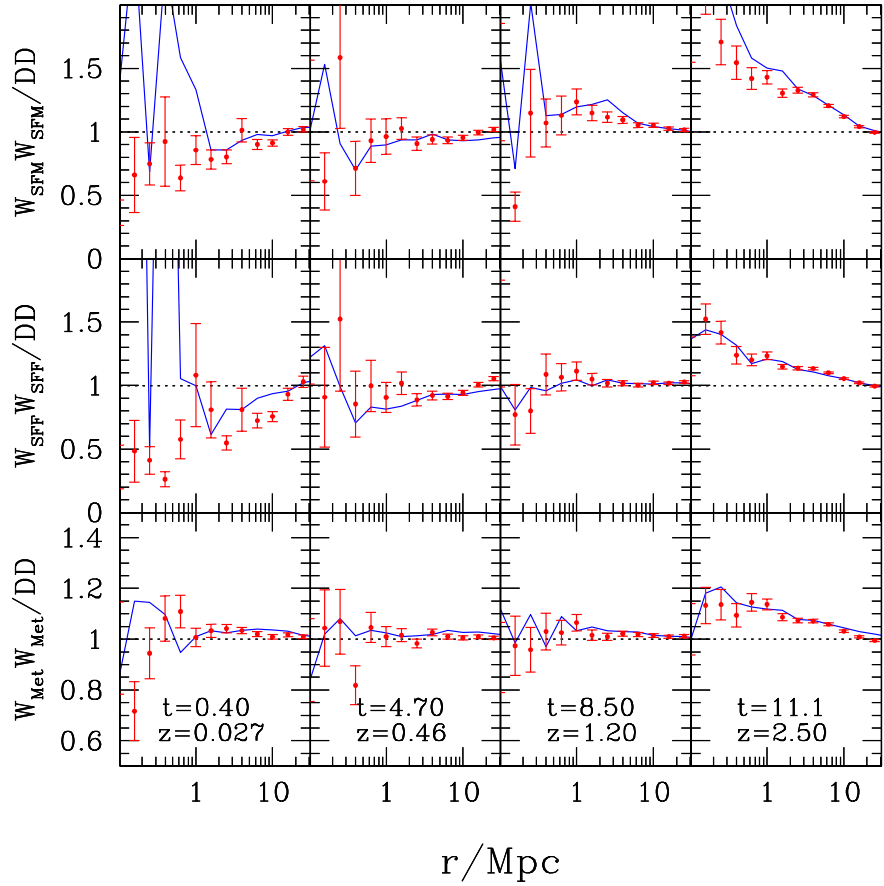}
 \caption{Mark correlation functions using weights associated 
          with different lookback times, $t$/Gyrs, or redshift, $z$.   
          Symbols with error bars show results for the more 
          luminous sample; lines without error bars represent 
          the fainter sample.  In both catalogs:  
          close pairs today had higher than average star formation 
          at $z=2.5$, average star formation at $z=1$, 
          and lower than average star formation more recently.  
          This is true whether one counts the total stellar mass 
          formed at $z$ (top panels), or the fraction of the 
          total stellar mass produced over the object's entire
          history which was formed at $z$ (middle panels).   
          Bottom panels show that close pairs which formed their 
          stars 11~Gyrs ago tend to have metallicities which are 
          above average for that epoch, but that at smaller lookback 
          times there is little correlation between metallicity and 
          present day environment. Figure from \cite{mark}.}
 \label{fig:xiSFM}
\end{figure}

The top panel of Figure~\ref{fig:xiSFM} shows a mark correlation 
analysis \footnote{Because it is impossible to define what a "cluster" or a "filament" is, it is more useful to compute the correlation properties of the, e.g., the star formation history, luminosity etc... the maks. This is what a mark correlation provides} of the total stellar mass formed at $z$, for a few 
lookback times.  It indicates that, compared to the average 
galaxy, objects with close pairs today were forming more stellar 
mass 11~Gyrs ago, but that they have been forming stellar mass at 
below average rates more recently.  Some of this trend is due to 
the fact that close pairs tend to be more luminous, so they are likely to contain 
more stellar mass.  To remove this effect, we have divided the 
stellar mass formed by an object at $z$ by the total stellar mass 
it ever formed.  The middle panels show the result of using 
this fraction of stars formed at $z$ as the weight; close pairs 
today had larger than average star formation fractions 11 Gyrs ago, 
average star formation fractions at redshifts of order unity, and 
smaller than average star formation fractions more recently.  
Thus, our analysis provides graphic evidence that the objects which 
formed most of their stellar mass at high redshifts are currently 
in clusters, where the current star formation rate is smaller than 
average.  The anti-correlation between star formation (mass or 
fraction) and environment persists up to lookback times of 5~Gyrs.

The bottom panels show a similar analysis when metallicity, 
$Z/Z_\odot$, is the mark.  The close pairs which had above average 
star formation fractions at $z=2.5$ also tend to have above 
average metallicities.  There are no clear correlations with 
environment in the other panels.  
Thus, these measurements \cite{mark}  indicate that the stellar populations of the 
most massive halos are old and metal rich.  The population of objects 
which had above average star formation fractions and metallicities 
at large lookback times are over-represented in clusters today.  

\begin{figure}[!htp]
\includegraphics[width=\columnwidth]{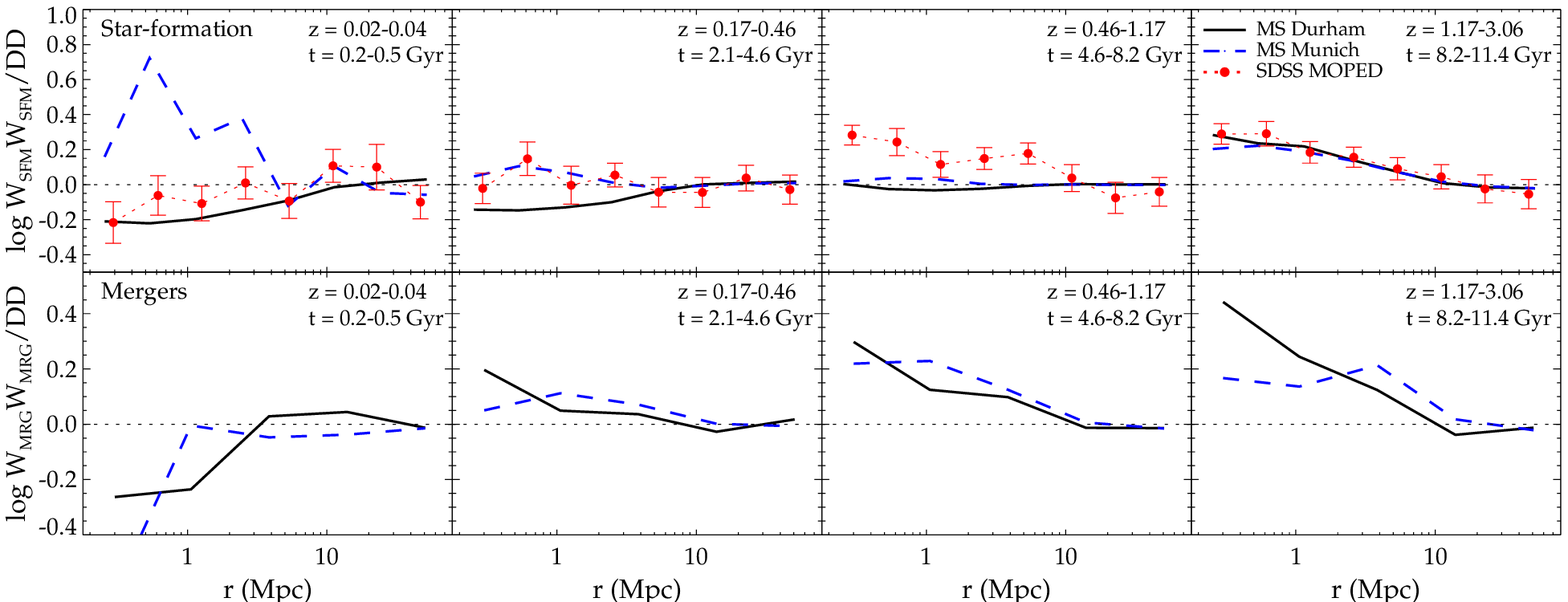}
\caption{Top panels: Evolution of the weighted correlation functions as defined by the total stellar mass formed at different lookback time or redshift bins for the Millennium Simulation using the results from Durham model (solid lines) and Munich model (dashed lines), and for the observational results from the SDSS data using the MOPED (dotted lines; filled circles) approach to retrieve the Star Formation History of galaxies.  Bottom panels: Millennium Simulation merger analysis; the weights are associated to the number of major mergers (with mass ratio larger than $1/3$) that the main progenitor of a $z=0$ galaxy experienced at each redshift. Figure from \cite{Abilio}.}
\label{fig:3}
\end{figure}

On the observational side, the recent advances in the field of spectral synthesis make it possible to retrieve detailed star formation histories from the spectra of nearby galaxies. Therefore, the evolution of mass assembly in galaxies through star formation as predicted by semi-analytical models can be properly compared with results from observational analysis. The mark correlation analysis can be then used to investigate the scale dependence of the star formation history of galaxies as predicted by the models and observations. In Fig.~\ref{fig:3}, we show results for semi-analytical models of galaxy formation from Durham and Munich and the SDSS data. In the top panels of this figure, the mark analysis were done considering the total stellar mass formed (SFM) at a given lookback time bin as the marks.  The most interesting result shown in Fig.~\ref{fig:3} is the excellent agreement between the mark correlations obtained by the semi-analytical models, especially the Durham model, and the observational data. The overall behaviour of the correlations in all lookback time bins shown in this figure indicates that close galaxy pairs as seen at $z=0$ formed more stellar mass at $\sim 10$~Gyr than the average, while more recently this trend is the opposite, with close pairs showing low levels of SF activity in comparison to the average. Note that for the first time bin, the Munich model predicts a positive correlation at smaller scales which is not caused by noise in the data, since the counts are similar to the Durham model. A possible explanation for these trends can be seen in the bottom panels of Fig.~\ref{fig:3}, where we show the mark correlations for the models considering the number of major mergers experienced by a galaxy in a given redshift bin. There is a clear relation between the increment in the star formation activity of close  $z=0$ galaxy pairs and the excess of major mergers at $z > 1$, thus reflecting that mergers played a significant role in the build up of galaxies in the high-z Universe.

Thus mergers seem to be the driver of shaping the star formation history through its effect in building the mass assembly of galaxies over time and as a function of environment. It is in this sense that mass alone determines the evolution of a galaxy. 

\section{How should we classify galaxies?}

We have seen that massive galaxies today form their stars early and less massive galaxies today form their stars late, thus present stellar mass alone is enough to predict the star formation history of a galaxy. Also, a mark correlation analysis of SDSS galaxies using MOPED-derived 
ages, metallicities and star formation histories shows that close 
pairs tend to host stellar populations which are older than average.  Close pairs also tend to have formed a larger fraction of their stars at $z\approx 3$ than average, but 
the star forming fraction at $z<1$ of such pairs is below average.  The objects which were forming stars at $z\approx 3$ have above average metallicities. These trends do not depend significantly on the mean luminosity of the sample, so they are approximately independent of stellar mass.  

Close pairs are dominated by galaxies in massive halos.  Hence, 
the MOPED results indicate that galaxies in clusters today are forming 
stars at below average rates, whereas they formed stars at above 
average rates at $z>1$.  Since clusters formed from overdense 
regions in the early Universe, the MOPED  results imply that cosmic star 
formation has moved from dense to ever less dense regions.  

What is the role of spin in the evolution of galaxies? \cite{Zach} used an indirect method to compute the dark matter spin of galaxies and combined it with MOPED determinations of the star formation history of the SDSS galaxies.   Exploiting the large sample available, they studied the influence of dark matter spin on galaxy mass, star formation history, and environment.  They found that galaxy dark matter spin and stellar mass are anti-correlated: lower stellar mass galaxies exhibit broader and generally higher distribution of spins than high-mass galaxies. Furthermore, according to halo mass estimates determined from the MOPED stellar masses, galaxies which have formed $1-6\%$ of their stellar mass in the past $0.2$ Gyr also have typically broader and higher-$\lambda$ spin distributions than galaxies that have formed a large fraction of their stellar mass at look-back times larger than $10$ Gyr. Although mass is the prime parameter determining the current star formation rate, the galaxy spin parameter might play a weak secondary role, with higher-spin galaxies having more current star formation at given stellar mass. \cite{Zach} also looked at environmental effects: using the RMF catalogue of galaxy clusters in the SDSS they found a very weak anti-correlation between the value of dark matter spin and proximity to a cluster, but such as would be consistent with mass and star formation being positively correlated with cluster proximity (see \citet{Abilio} for a comparison of the environmental dependence of galaxies with numerical models of galaxy formation). A marked correlation study also shows no strong correlation with galaxy separation, in agreement with previous studies' predictions.

Therefore, spin plays only a secondary role in shaping the evolution of a galaxy. From the analysis presented above it seems that we could classify galaxies as follows: use the present stellar mass to classify galaxies in different star formation and metallicity history bins. Next, add the environment information by classifying galaxies according to how close they are to their neighbors: that will produce a record of how was its past dark halo merger history. Finally, record the dark halo spin as a final, albeit secondary effect, shaping the evolution of  galaxy.

\end{document}